\documentclass[showpacs,floatfix,preprintnumbers,amsmath,amssymb,12pt]{revtex4}

\usepackage{graphicx}
\usepackage{epsfig}
\usepackage{dcolumn}
\usepackage{bm}

\begin{document}

\title{Isospin-dependent clusterization of Neutron-Star Matter}

\author{C. Ducoin$^{(1,2)}$, Ph. Chomaz$^{(1)}$  and F. Gulminelli$^{(2)}$\footnote{member of the Institut Universitaire de France}}
\affiliation
	{
	(1) GANIL (DSM-CEA/IN2P3-CNRS), B.P.5027, F-14076 Caen c\'{e}dex 5, France \\
	(2) LPC (IN2P3-CNRS/Ensicaen et Universit\'{e}), F-14050 Caen c\'{e}dex, France
	}

\begin{abstract}
Because of the presence of a liquid-gas phase transition in nuclear matter,
compact-star matter can present a region of instability against the formation of clusters.
We investigate this phase separation in a matter composed of neutrons, protons and electrons, within a Skyrme-Lyon mean-field approach.
Matter instability and phase properties are characterized through the study 
of the free-energy curvature. 
The effect of $\beta$-equilibrium is also analyzed 
in detail, and we show that the opacity to neutrinos has an influence on the presence of clusterized matter in finite-temperature proto-neutron stars. 
\end{abstract}


\maketitle

\section{Introduction}

Nuclear matter is known to present a phase transition of the liquid-gas type,
which occurs at sub-saturation density up to a critical temperature
\cite{Bertsch-PLB126,Finn-PRL49,Das-PhysRep}.
In laboratory experiments, this phase diagram can be explored 
only through heavy-ion collisions; in particular
the multi-fragmentation phenomenon, 
observed for collisions around the Fermi energy and above,
has been interpreted as the occurrence 
of the nuclear liquid-gas phase transition
rounded by finite-size effects 
\cite{Borderie-NPA734,FG-HDR,Bondorf-PhysRep,Rivet-NPA749,Trautmann-NPA-752}.
From the theoretical point of view, 
the isospin properties of the phase diagram have been studied 
extensively\cite{Muller-Serot-PRC52,Baran-PhysRep,CD-A1}
and we know that the phase transition concerns isospin-symmetric ($\rho_n = \rho_p$) 
as well as asymmetric ($\rho_n \neq \rho_p$)
matter. Then the transition phenomenology may also affect
the physics of compact stars,
which are made of dense, neutron-rich matter
\cite{Glendenning-PhysRep,Lattimer-PhysRep}.
In particular neutron-star properties may 
be influenced by the presence of exotic "pasta phases" in their inner crust
\cite{Ravenhall-PRL50,Lassaut-AA183,Watanabe-PRC68,Horowitz-PRC69}.
Density fluctuations due to the liquid-gas instability may also play a role 
in the dynamics of supernova explosions, by affecting neutrino propagation
\cite{Buras-PRL90,Margueron-PRC70}.

\begin{figure}
\begin{center}
\includegraphics[width=0.5\linewidth]{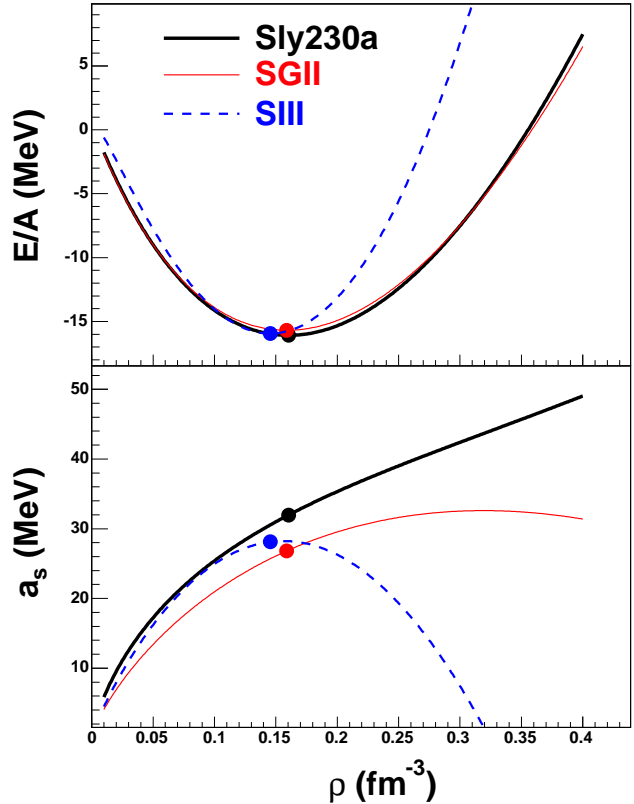}
\caption
	{
	Energy proporties at zero temperature for the different Skyrme forces used in this paper.
	Saturation points are indicated by dots.
	Top : energy per nucleon of symmetric matter. 
	Bottom : density dependence of the asymmetry coefficient 
	$a_s=\frac{1}{2} \left.\frac{\partial^2E/A}{\partial I^2}\right|_{I=0}$, with $I=\rho_3/\rho$.
	}
\label{FIG:Skyrme}
\end{center}
\end{figure}

In this paper, we study the instability of star matter against the liquid-gas phase transition.
At the densities of interest for our study, it can be considered as a medium composed of neutrons, protons and electrons, under the constraint of strict electroneutrality.
We will describe nuclear matter (neutrons and protons, denoted "$NM$")
in a mean-field approach, using different Skyrme effective interactions. 
The Skyrme-Lyon parameterization Sly230a, already employed for astrophysical
applications in ref.\cite{Douchin-PLB485}, is taken as a reference, 
since it has been especially fitted to describe neutron-rich matter
\cite{Chabanat-NPA627,Douchin-NPA665}. 
It is compared to two other Skyrme forces :
the earlier SGII\cite{Giai-NPA371}, whose parameters are optimized
to describe nuclear collective motion,
and SIII, one of the first and more simple Skyrme parameterizations
\cite{Beiner-NPA238}.
Fig. \ref{FIG:Skyrme} shows the density behavior of these different forces
in the isoscalar (upper part) as well as isovector (lower part) channel.

We can see that realistic interactions, fitted on measured nuclear properties
as Sly230a and SGII, are perfectly superposed for symmetric nuclear matter, while SIII presents an abnormally large incompressibility, given by the curvature at saturation :	$K_{\infty}=9\rho_0^2 \left.\frac{\partial^2 E/A}{\partial \rho^2}\right|_{\rho=\rho_0}$. The density dependence of the symmetry 
energy is less well constrained by experimental data\cite{Baran-PhysRep},
and shows very different behaviors with the different interactions.

Electrons are described as a Fermi gas, whose density is set equal to proton density.
Nuclear matter with electrons (representing star matter) will be denoted "$NMe$".

In section
\ref{SEC:homog},
we present nuclear-matter thermodynamics and compare the 
spinodal-instability region for $NM$ and $NMe$.
In section \ref{SEC:cluster},
we consider the instability of homogeneous $NMe$
against finite-size density fluctuations, leading to matter clusterization.
Some qualitative features of the clusters are deduced
from the  free-energy curvature properties of homogeneous matter in the 
framework of a spinodal-decomposition scenario,
where cluster characteristics are determined by the dominant fluctuation mode 
at the onset of phase separation.
Finally, in section \ref{beta} 
we investigate the effect of neutrinos on star-matter instability
when $\beta$-equilibrium is assumed.

\section{Global instability of homogeneous matter} \label{SEC:homog}

In this section, we establish the 
thermodynamic instability properties of infinite and homogeneous matter.
For a fixed temperature, the free-energy curvature in the density plane $(\rho_n,\rho_p)$
determines the spinodal region. Inside the spinodal, a thermodynamic fluctuation 
of the homogeneous system can lead to a reduction of the free energy :
this fluctuation is then amplified, leading to phase separation. 
Since the resulting phases are both homogeneous and infinite, any
even infinitesimal charge density would lead to a divergent Coulomb energy:
the thermodynamic limit thus imposes that each phase is exactly neutral
\cite{PC-europhys}.

In the actual cooling of a proto-neutron star, the characteristic 
time scales may be long enough respect to the typical relaxation
times associated with the strong interaction, for the system to be 
in thermodynamic equilibrium at all steps of its evolution.
If this is the case, the spinodal zone is not directly pertinent
for star-matter thermodynamics. However the spinodal region 
contains points of locally inversed curvature, 
and is always by construction included inside the 
region of globally inversed curvature, namely the coexistence zone
\cite{Muller-Serot-PRC52,CD-A1}.
Therefore our results on the thermodynamic instabilities will keep
being relevant for the equilibrium mixed phase.

We will first evaluate the spinodal region in nuclear matter, in the 
absence of Coulomb interaction.
Then, electroneutrality will be insured by the presence of electrons, 
which results in a strong quenching of instability.

\subsection{Nuclear-matter spinodal region}\label{SUBSEC:homog-spino}

Nuclear matter involves an infinite quantity of nucleons interacting through the nuclear force.
The Coulomb interaction is not taken into account in this model, 
but neutrons and protons are distinguished by their isospin. 

In a mean-field approach, each nucleon is an independent particle evolving in a potential 
resulting from the averaged effect of other nucleons. 
Let us denote $\hat H$ the hamiltonian of the system, and $\hat \rho_q$ the density matrix of each nucleon species, the index $q$ representing neutrons ($q=n$) or protons ($q=p$). 
The mean field $\hat{W}_{q}$ is defined by the relation $\delta \langle \hat{H}\rangle =Tr(\hat{W}_{q}\delta \hat{\rho_{q}})$. 
Within this approximation, the average energy density of homogeneous nuclear matter 
$\mathcal H^h = \langle \hat{H}\rangle/V$ is a functional of one-body 
densities only: 
particle densities $\rho _{q}$ and kinetic densities $\tau _{q}=\langle \hat{p}^2\rangle_q/\hbar^2$.
It is useful to introduce isoscalar and isovector densities :
\begin{equation}
\begin{array}{ll}
\rho =\rho _{n}+\rho _{p}\;, & \;\tau =\tau _{n}+\tau _{p}, \label{EQ:rho} \\
\rho _{3}=\rho _{n}-\rho _{p}\;, & \;\tau _{3}=\tau _{n}-\tau _{p}.
\end{array}
\end{equation}
Using a Skyrme effective interaction, the energy density of homogeneous, spin-saturated matter with no
Coulomb interaction, can be written as :
%
%
\begin{equation}
\mathcal{H}=\mathcal{K}+\mathcal{H}_{0}+\mathcal{H}_{3}+\mathcal{H}_{eff} \label{EQ:E_HNM}
\end{equation}
In this expression, $\mathcal{K}$ is the kinetic-energy term, 
$\mathcal{H}_{0}$ a density-independent two-body term, $\mathcal{H}_{3}$ 
a density-dependent term, and $\mathcal{H}_{eff}$ a momentum-dependent term:
\begin{equation}
\begin{array}{lll}
\mathcal{K}&=&\frac{\hbar ^{2}}{2m}\tau\\ 
\mathcal{H}_{0} &=&C_{0}\rho ^{2}+D_{0}\rho _{3}^{2}\\
\mathcal{H}_{3} &=&C_{3}\rho ^{\sigma +2}+D_{3}\rho ^{\sigma }\rho _{3}^{2}\\
\mathcal{H}_{eff} &=&C_{eff}\rho \tau +D_{eff}\rho _{3}\tau _{3}
\end{array}
\end{equation}
%
%
where the coefficients $C_i$ an $D_i$ are linear combinations 
of the standard Skyrme parameters \cite{Chabanat-NPA627,CD-A1}.
The mean field is given by the expression :
\begin{equation}
\hat{W}_{q}=\frac{\partial \mathcal H^h}{\partial \tau _{q}}
\frac{\hat{p}^{2}}{\hbar ^{2}}+\frac{\partial \mathcal H^h}{\partial \rho _{q}} 
=\frac{1}{2m_{q}^{*}}\hat{p}^{2}+U_{q},
\label{EQ:MeanField}
\end{equation}
where $m^{*}_{q}$ is the effective mass of particles of type $q$. The individual energy levels read $\epsilon_{q}^{i}=\frac{p_{i}^{2}}{2m_{q}^{*}}+U_{q}$. In the grand-canonical formalism, they are occupied according to Fermi-Dirac statistics with occupation numbers:
\begin{equation}
n_{q}(p)=1/ \left[ 1+exp(\beta (p^{2}/2m_{q}^{*}+U_q-\mu _{q})) \right],
\label{EQ:distribution}
\end{equation}
where $\beta$ is linked to the temperature by $\beta=1/T$, and $\mu_q$ is the chemical potential of particles of type $q$.
All one-body densities are then given by Fermi integrals and
the mean-field grand-canonical partition sum can be expressed 
in the same way as for an ideal gas:
\begin{equation}
\frac{\ln Z_{0}}{V}=2 \sum_q \int_{0}^{\infty }
\ln(1+e^{-\beta (\frac{p^{2}}{2m_{q}^{*}}-\mu _{q}^{\prime })})
\frac{4\pi p^{2}}{h^{3}} dp
=\sum_q \frac{\hbar^2}{3m_{q}^{*}}\beta \tau _{q},
\end{equation}
where $\mu _{q}^{\prime }=\mu_q-U_q$.

For fixed values of $\rho_n$ and $\rho_p$ at a given temperature, 
the value of chemical potentials as well as kinetic densities are obtained
by iteratively solving the self-consistent relations between $\rho_q$, $\mu_q$ and $m_{q}^{*}$. 
We can then calculate the entropy as a function of densities. 
Indeed the mean-field approach approximates the real entropy $S$ 
by the corresponding mean-field entropy $S^{0}$\cite{Vautherin-ANP22} :
\begin{equation} 
s=S/V \simeq S_{0}/V=\ln Z_{0}/V+\beta(\langle\hat{W}\rangle_{0}/V-\mu_{n}\rho_n-\mu_{p}\rho_p).
\end{equation}
%
 
For a given temperature $T=\beta^{-1}$, statistical equilibrium corresponds
to the minimization of the free energy $F=Vf$, related to the entropy $S$ by a Legendre transform.
Per unit volume, this relation is
$f=\mathcal H^h-s/\beta$ and $f$ is a function of nucleon densities $\rho_n$ and $\rho_p$.
If the free-energy surface presents a concavity, it can be minimized by linear interpolation
(phase mixing). 
The equilibrium configuration corresponds to a phase coexistence determined by Gibbs contruction, 
which is a several-dimension generalization of the well-known Maxwell construction
\cite{Muller-Serot-PRC52}.

We now turn to spinodal instabiliy. 
Let us consider an homogeneous system submitted to a thermodynamic fluctuation,
consisting in a separation into two portions distinguished by an infinitesimal density shift. 
Both portions correspond to infinite and homogeneous systems.
If the fluctuation leads to a reduction of the global free energy, 
it will be amplified in the successive time evolution:
the initial system is locally unstable.
The spinodal region is then defined  as the ensemble of unstable points, 
\emph{i.e.} points for which the free-energy surface 
is convex in at least one direction.
This is determined by the curvature matrix :
\begin{equation}
C^h_{NM}=
\left( 
\begin{array}{ll}
\partial^2 f/ \partial \rho _n^2 & \partial^2 f/ \partial \rho _n\partial \rho _p\\ 
\partial^2 f/ \partial \rho _p \partial \rho_n & \partial^2 f/ \partial \rho _p^2\\ 
\end{array}
\right)
=
\left( 
\begin{array}{ll}
\partial\mu _{n}/\partial\rho _{n} & \partial\mu _{n}/\partial\rho _{p}\\ 
\partial\mu _{p}/\partial\rho _{n} & \partial\mu _{p}/\partial\rho _{p}\\ 
\end{array}
\right)
\end{equation}
This matrix is defined for each point $(\rho_n,\rho_p)$,
and its lower eigen-value $C_{<}$ gives the minimal free-energy curvature in this point.
The eigen-vector $\mathbf{u}_<$ associated with the minimal curvature, 
named "instability direction", gives the direction of phase separation.

\begin{figure}
\begin{center}
\includegraphics[width=1\linewidth]{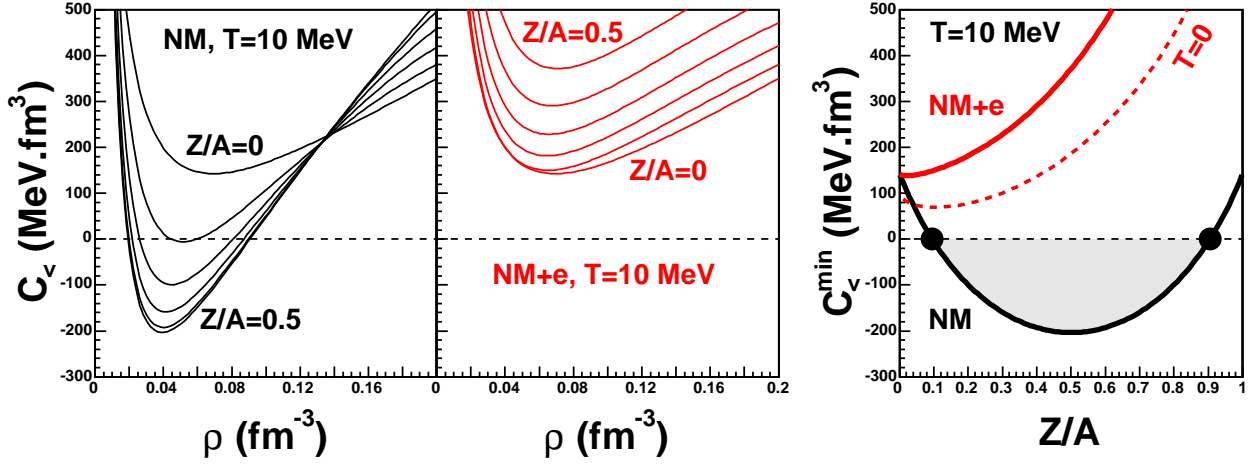}
\caption
	{
	Minimal free-energy curvature obtained with the Sly230a interaction.  
	Ordinary nuclear matter ($NM$)presents a spinodal region,
	which is suppressed when an electron gas is added to neutralize the 
	proton charge ($NMe$).
	Left : lower eigen-value as a function of the total density $\rho$, 
	for different $Z/A$ from $Z/A=0.5$ to $Z/A=0$, equally spaced of $0.1$.
	Right : minimal curvature as a function of $Z/A$. 
	Dotted line : $MNe$ at $T=0$.
	}
\label{FIG:spinosup}
\end{center}
\end{figure}

We have represented on the left part of fig.\ref{FIG:spinosup} the minimal curvature $C_<$ 
as a function of total density, for different values of the proton fraction $Z/A$, 
at a fixed temperature of 10 MeV : some of the curves have a negative region. 
On the right part, the minimal value of $C_<$ obtained for each asymmetry 
is represented as a function of $Z/A$. 
This shows that, for a finite range of asymmetry, homogeneous matter presents a spinodal instability, 
\emph{i.e.} a negative curvature of the free energy. 
This is observed until a critical temperature of about $14$ MeV.

\subsection{Electron quenching of instability} \label{SUBSEC:homog-equench}

Let us now take the electric charge into account.
Matter must be strictly neutral at a macroscopic scale, 
in order to avoid the divergence of Coulomb energy.
In this paper, we consider star matter where the proton charge 
is neutralized by a Fermi gas of electrons.
Studying thermodynamic fluctuations,
each portion in which the system separates is submitted to the constraint of strict electroneutrality : 
the electron density is then everywhere exactly equal 
to the proton density, $\rho_e=\rho_p$.
In this case, electrons do not add any degree of freedom, and 
the free-energy curvature matrix remains two-dimensional : 
\begin{equation}
C^h_{NMe}=
\left( 
\begin{array}{ll}
\partial\mu _{n}/\partial\rho _{n} & \partial\mu _{n}/\partial\rho _{p}\\ 
\partial\mu _{p}/\partial\rho _{n} & \partial\mu _{p}/\partial\rho _{p}+\partial\mu _{e}/\partial\rho _{e}\\ 
\end{array}
\right)
\end{equation}
The presence of electrons contributes to this matrix via the electron susceptibility $\chi_e^{-1}=\partial\mu _{e}/\partial\rho _{e}$.
This term is readily evaluated in the approximation
of a degenerated, ultrarelativistic Fermi gas.
Because of the low electron mass, 
these approximations are valid until very low densities.
For a degenerated Fermi gas, 
the Fermi momentum $p_{F_e}$ is given by the relation : 
\begin{equation}
\rho_e=\rho_p=\frac{(p_{F_e} c)^3}{3\pi^2(\hbar c)^3}.
\end{equation}
In the ultra-relativistic approximation, 
the chemical potential is equal to the Fermi momentum $\mu_e=p_{F_e} c$, 
so it is related to the density by :
\begin{equation}
\mu_e = \hbar c (3 \pi^2 \rho_e)^{1/3},
\end{equation}
which gives the electron contribution to the curvature matrix :
\begin{equation}
\partial \mu_e/\partial \rho_e 
= \chi_e^{-1}
= \hbar c (\frac{\pi^2}{9 \rho_e^2})^{1/3.}
\end{equation}

In the density region of interest for our study, the numerical value of
$\chi_e^{-1}$ is large in comparison with the other terms of the free-energy curvature matrix
\footnote{  
For example for symmetric matter at saturation density, the susceptibility $\chi_e^{-1}$ is  $1052.45 MeV$, while the nuclear free-energy second derivatives ($\partial\mu _{q}/\partial\rho _{q}$ and $\partial\mu _{n}/\partial\rho _{p}$) are respectively $570.864$ and $-202.96 MeV$ at zero temperature ($588.407$ and $-202.96 MeV$ at a temperature of $10 MeV$).
}. 
As a result, adding an electron gas leads to 
a drastic effect on the eigen-modes of the matrix. 
Because of its positive sign,
it leads to a quenching of the instability.
This is illustrated on fig.\ref{FIG:spinosup} : 
the middle part presents the minimal curvature as a function of total density for $MNe$ 
with different values of $Z/A$, at $T=10 MeV$. 
The lowest minimal curvatures are reported on the right part 
as a function of $Z/A$. No negative value of the curvature is found,
and the same is true at zero temperature.
Since the extension of the spinodal region monotonically increases 
with decreasing temperatures, this result implies that $NMe$
does not present any thermodynamic instability. 

\begin{figure}
\begin{center}
\includegraphics[width=0.8\linewidth]{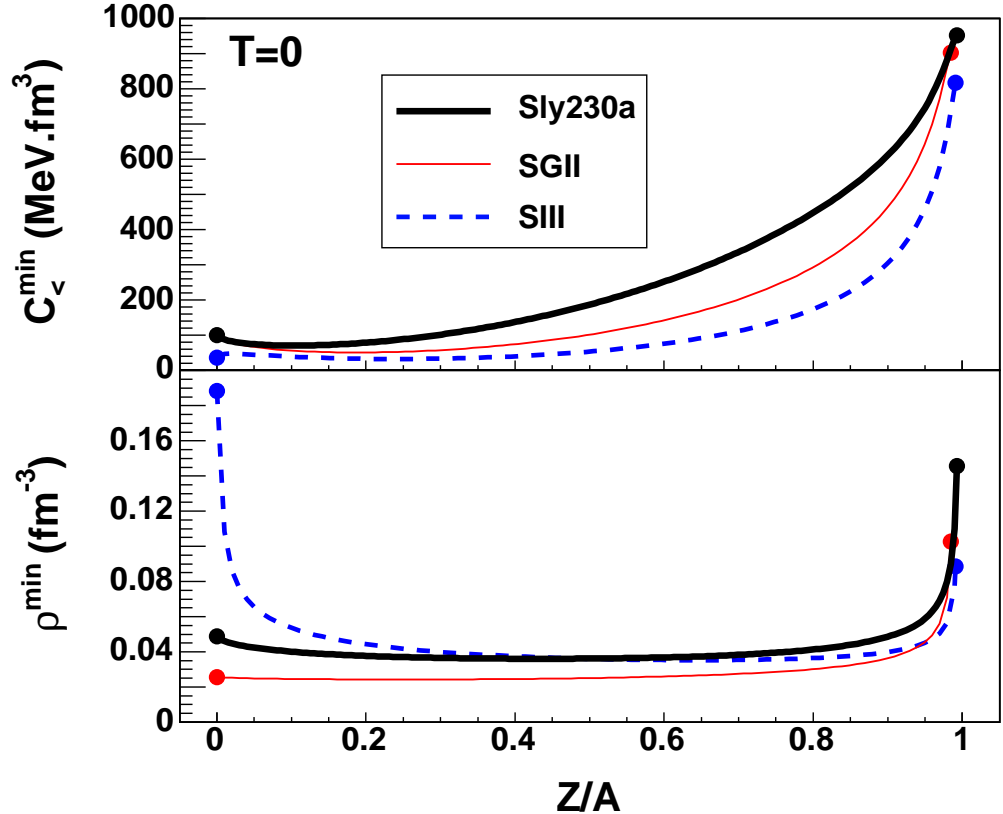}
\caption
	{
	Suppression of spinodal instability in nuclear matter with electrons,
	for different Skyrme forces at $T=0$.
	Top : minimal free-energy curvature obtained for a given proton fraction, as a function of $Z/A$.
	Bottom : density at which this minimum is obtained.
	}
\label{FIG:spinosup-3F}
\end{center}
\end{figure}

%
%
%
The value of $C^{min}_<$ depends on the details of the nuclear interaction
used for the calculation of neutron and proton chemical potentials.
For one given point, the sign of $C_<$ is given by 
$\det C^h_{NMe}=\det C^h_{NM}+\partial_{\rho_n} \mu_n*\partial_{\rho_e}\mu_e$. 
It can be negative if $\det C^h_{NM}$ has a sufficiently strong negative value,
\emph{ie} nuclear-matter thermodynamic instability is strong enough to balance the positive term due to the electron gas. 
This is does not happen for any of the three Skyrme forces we have considered,
as shown by figure \ref{FIG:spinosup-3F} : 
the instability is totally suppressed by electron incompressibility.
However, this result is model dependent. For example, in reference \cite{Providencia-PRC73}
the spinodal instability is studied in a relativistic-mean-field approach, using the NL3 set of parameters :
the authors obtain a small thermodynamic instability region in the case of nuclear matter with electrons.
This region, located at low density, is very reduced with respect to nuclear-matter spinodal,
as expected from electron quenching.
%
%

\section{Clusterization of $NMe$}  \label{SEC:cluster}

In the previous section we have shown that the electron
gas in stellar matter, because of its high incompressibility, 
acts against the formation of macroscopic charge dishomogeneities,
thus completely quenching the liquid-gas phase transition,
at least in the mean-field approximation adopted here.
This does not mean that we should expect stellar matter 
at subsaturation density to be homogeneous.
Indeed it has been recognized since a long time\cite{Negele} 
that the ground state of neutron-rich stellar matter at low density
should be a Coulomb lattice of exotic nuclei embedded in a neutron 
gas, meaning that we should expect a solid-liquid transition 
between the core and the inner crust of a neutron star
\footnote{It is worthwile mentioning that, contrary to ordinary condensed 
matter, this solid-liquid transition is a continuous transition
\cite{PC-europhys}}.
Numerical simulations performed with (semi)-classical models 
\cite{Watanabe-PRC68,Horowitz-PRC69} indicate that such dishomogeneous 
phases should persist at finite temperature and in a large
range of asymmetry.
%
%
Clusterisation of hot stellar matter is also considered in a statistical approach in 
ref. \cite{Botvina-PLB584,Botvina-PRC72}.
%
%
A way to address the possible existence of a phase of clusterized 
matter in mean-field approaches, is to study the instability of nuclear matter with electrons against 
finite-size density fluctuations\cite{Pethick-NPA584}.
If such fluctuations lead to a reduction of the free-energy, 
they will be amplified through the process of spinodal decomposition\cite{PC-PhysRep}
until the formation of structures of alternatively high and low density. 
Homogeneous matter is then unstable against clusterization. 
Clusters correspond to the high-density regions, which can be seen as 
liquid nuclear drops immersed in a nuclear gas.
Considering the short time scales associated with spinodal decomposition, 
the cluster characteristics 
are determined by the free-energy curvature properties at the onset of instability, when density fluctuations have small amplitudes\cite{PC-PhysRep}.


\subsection{Free-energy variation with finite-size fluctuations}

The instability region of homogeneous $NMe$ against clusterization
is determined by studying how the system free energy changes,
when a finite-size density fluctuation is introduced. 

If fluctuations occur on a finite microscopic scale,
electron and proton density can fluctuate independently : 
only their mean values are constrained to be equal, insuring macroscopic electroneutrality.
Fluctuations thus affect independently the three species of the medium
(neutrons, protons and electrons), whose densities become :
\begin{equation}
\rho_q = \rho^0_q + \delta \rho_q,
\end{equation}
with $q \in \{n,p,e\}$.
Each density variation can be expressed by a Fourier transform
\begin{equation}
\delta \rho_q = \int d \vec k a_q(\vec k) e^{i\vec k \cdot \vec r},
\end{equation}
with $a_q(\vec k)=a^*_q(-\vec k)$ to insure that $\delta \rho_q$ is real.
Since the different wave vectors $\vec k$ are decoupled in the global 
free-energy variation, the problem reduces to the 
study of plane-wave density fluctuations:
\begin{equation}
\delta \rho_q = A_q e^{i\vec k \cdot \vec r} + c.c.,
\end{equation}
where each species is associated with a different amplitude.
Thermodynamic fluctuations, studied in the previous section, can be 
obtained as the limit $k \rightarrow 0$ of this model.

To evaluate the free-energy variation, we consider 
a Thomas-Fermi approximation, \emph{i.e.} the density variation is supposed smooth enough to allow in each point the definition of a Fermi sphere corresponding to the local density.
Then, in each point of density $ \rho_q(\mathbf{r})=\rho^0_q+\delta \rho_q $, 
the local bulk term of the free energy is equal to the free energy $f^h$ 
of an infinite homogeneous system at the same density. 
The global bulk free energy of the system is the space average of this local term :
\begin{equation}
f^b=\frac{1}{V} \int {f^h    (  \{\rho_q(\mathbf{r})\}  )   d\mathbf{r}    }.
\end{equation}
%
In the small-amplitude limit, the 
integration leads to $f^b=f^h(\{\rho^0_q\})+\delta f^b$, with :
\begin{equation}
\label{EQ:delta-fb}
\delta f^b
=\sum_{i,j} \frac{A_{i}A^*_{j} +  A^*_{i}A_{j}}{2}  \left( \frac{\partial^2{f^h}}{\partial \rho_i \partial \rho_j} \right)_{\{\rho_q^0\}}
=\sum_{i,j} \frac{A_{i}A^*_{j} +  A^*_{i}A_{j}}{2}  \left( \frac{\partial{\mu_j}}{\partial \rho_j} \right)_{\{\rho_q^0\}},
\end{equation}
where $i,j \in \{n,p,e\}$. First-order terms have vanished in the integration because the average density variation is zero.

The variation of the entropy is contained in the bulk term, since entropy depends only on the local density 
\cite{Vautherin-ANP22}.
However, in the case of a finite wave number $k$, the energy density is modified by two additional terms
\cite{Pethick-NPA584}, 
arising from the density-gradient dependence of the nuclear force, 
and from the Coulomb interaction.  
Denoting these two contributions  $\delta \mathcal E^{\nabla}$ and $\delta \mathcal E^c$ respectively, 
the free-energy variation is :
\begin{equation}
\delta f = \delta f^b + \delta \mathcal E^{\nabla} + \delta \mathcal E^c.
\end{equation}

Let us now explicit the gradient and Coulomb contributions.
In the presence of density gradients the nuclear energy density has the form $\mathcal H = \mathcal H^h + \mathcal H^{\nabla}$,
where $\mathcal H^h$ is given by eq.(\ref{EQ:E_HNM}) and the density-gradient term $\mathcal H^{\nabla}$ is expressed as: 
\begin{equation}
\mathcal H^{\nabla}
=C_{nn}^{\nabla}(\nabla \rho_n)^2
+C_{pp}^{\nabla}(\nabla \rho_p)^2
+2C_{np}^{\nabla} \: \:  \nabla \rho_n \cdot \nabla \rho_p.
\end{equation}
Coefficients $C_{ij}^{\nabla}$ are combinations of the Skyrme parameters given for example in ref.\cite{Chabanat-NPA627}:
\begin{equation}
\begin{array}{l}
C_{nn}^{\nabla}=C_{pp}^{\nabla}
=\frac{3}{16} \left[ t_1(1-x_1) - t_2(1+x_2) \right],\\
C_{np}^{\nabla}=C_{pn}^{\nabla}
=\frac{1}{16} \left[ 3t_1(2+x_1) - t_2(2+x_2) \right].
\end{array}
\end{equation}
The global contribution of this term to the energy density is given by the 
space average: 
\begin{equation}
\label{EQ:delta-enabla}
\delta \mathcal E^{\nabla}=\frac{1}{V} \int {\mathcal H^{\nabla}(\vec r) d \vec r}
= k^2 \sum_{i,j} \left(A_iA_j^*+A_i^*A_j \right) C_{ij}^{\nabla},
\end{equation}
where $i,j \in \{n,p\}$.

The Coulomb contribution $\delta \mathcal H^c$ is due to the independent density fluctuations of charged particles $\{e,p\}$.
Denoting $q_i=\sqrt{4\pi \epsilon_0} e_i$ the electric charge of a particle of type $i$, this creates the charge distribution :
\begin{equation}
\label{EQ:drc-dri}
\rho_c(\vec r)=\sum_{i} e_i A_i e^{i\vec k \cdot \vec r} + c.c.
\end{equation}
%
%
%
The consequent Coulomb energy per unit volume is :
%
%
%
\begin{equation}
\label{EQ:delta-ec}
\delta \mathcal E_c
=\frac{1}{2V} \int{  \frac {\rho_c (\mathbf{r}_1) \rho_c (\mathbf{r}_2)}{\left| \mathbf{r}_2-\mathbf{r}_1 \right|}   d\mathbf{r}_1 d\mathbf{r}_2 }
=\frac{4\pi}{k^2} \sum_{i,j} \frac{A_iA_j^*+A_i^*A_j }{2} ,
\end{equation}
where $i,j \in \{e,p\}$.

Summing the contributions (\ref{EQ:delta-fb}), (\ref{EQ:delta-enabla}) and (\ref{EQ:delta-ec}),
we have the total free-energy variation, expressed to the second order in $A_q$ :
\begin{equation}
\label{EQ:delta-f-bis}
\delta f=\sum_{ij} 
\frac{A_iA_j^*+A_i^*A_j}{2}
\left[ 
\left( \frac{\partial \mu_i}{\partial \rho_j}\right)_{\{ \rho_q^0 \}}
+2C_{ij}^{\nabla}k^2
+\frac{4\pi e_i e_j}{k^2} 
\right],
\end{equation}
which can be written in a matrix form in the three-dimensional space of 
density fluctuations $\tilde{A}=\left ( A_n, A_p, A_e\right )$ : 
\begin{equation}
\delta f={\tilde{A}^*} \mathcal C^f \tilde{A},
\end{equation}
where
%
%
%
\begin{equation}
\label{EQ:Cf}
\mathcal C^f=
\left( 
\begin{array}{ccc}
\partial\mu _{n}/\partial\rho _{n} & \partial\mu _{n}/\partial\rho _{p} & 0\\ 
\partial\mu _{p}/\partial\rho _{n} & \partial\mu _{p}/\partial\rho _{p} & 0\\
0 & 0 & \partial\mu _{e}/\partial\rho _{e}\\ 
\end{array}
\right)
+
k^2
\left( 
\begin{array}{ccc}
2C_{nn}^{f} & 2C_{np}^{f} & 0\\ 
2C_{pn}^{f} & 2C_{pp}^{f} & 0\\
0 & 0 & 0\\ 
\end{array}
\right) 
+
\frac{4\pi^2e^2}{k^2}
\left( 
\begin{array}{ccc}
0 & 0 & 0\\ 
0 & 1 & -1\\
0 & -1 & 1\\ 
\end{array}
\right) 
\end{equation}
%
is the free-energy curvature matrix.
The bulk term of the curvature is conserved in this new expression, 
with the difference that the electron term now belongs to an independent space. 
The density-gradient part of the nuclear interaction adds a term proportional to $k^2$, while the Coulomb interaction induced by 
the plane-wave charge distribution adds a term inversely proportional to $k^2$.

\subsection{$k$-dependent instability}

\begin{figure}
\begin{center}
\includegraphics[width=0.8\linewidth]{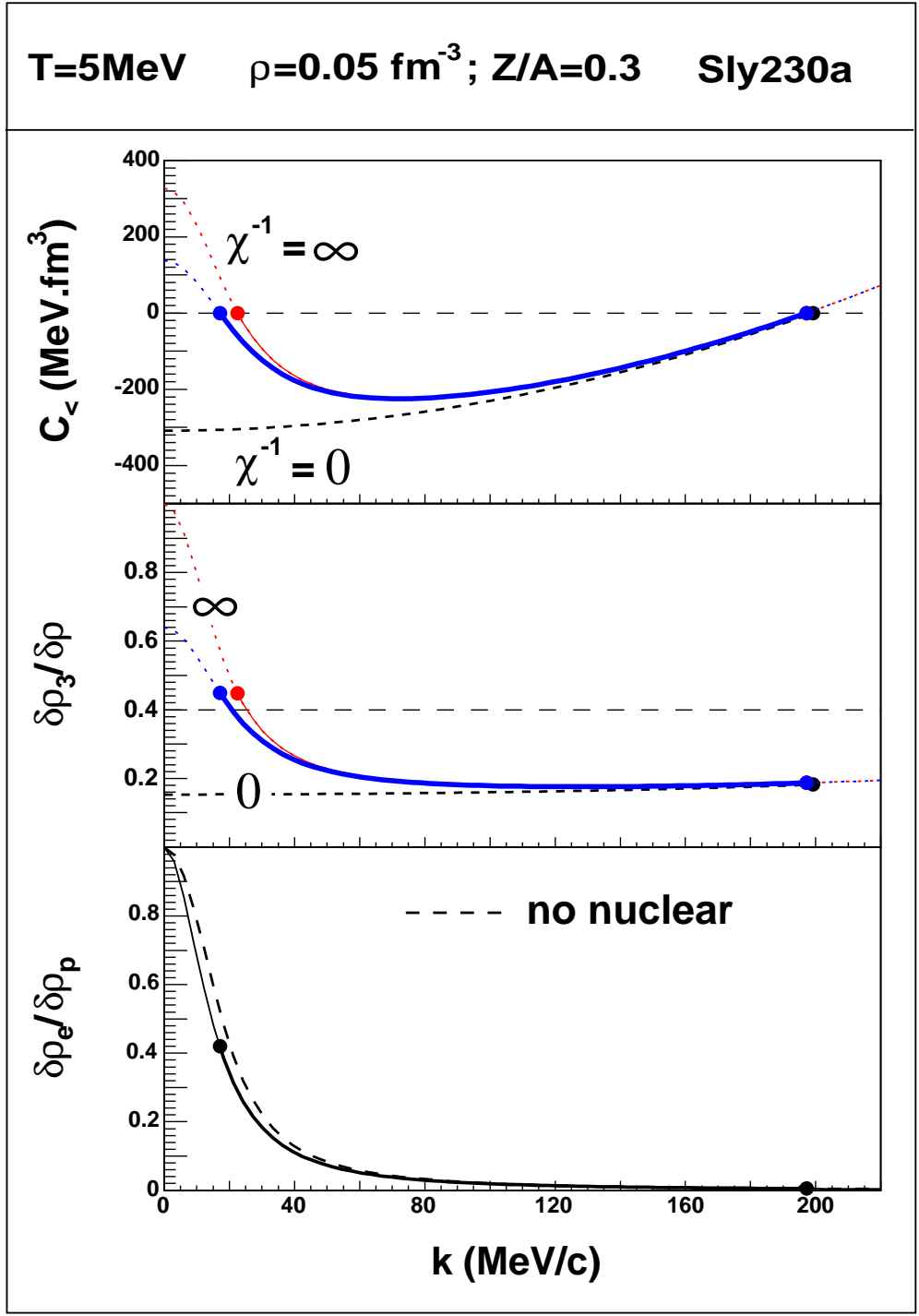}
\caption
	{
	Eigen-mode of the minimal free-energy curvature for $\rho=0.05 fm^{-3}$,
	$Z/A=0.3$, $T=5$ MeV,  
	as a function of the wave number $k$, calculated with the Sly230a Skyrme   
  interaction.
	Top : eigen-value.
	Middle : associated eigen-vector in the nuclear-density plane. 
	The $NMe$-curves are compared to two limiting cases 
	corresponding to a zero and infinite incompressibility of the negatively charged
	 gas.
	When the curvature is negative (full lines) this eigen-vector gives the
	 phase-separation 
	 direction. The dots give the points of zero curvature.
	The dashed line gives $\rho_3/\rho$ (direction of constant $Z/A$).
	Bottom : same as the middle part, in the $(\rho_p,\rho_e)$-plane.
	The dashed line gives $\delta \rho_e/\delta \rho_p$ when the nuclear force is zero.
	}
\label{FIG:Cf-point}
\end{center}
\end{figure}

An instability against matter clusterization 
corresponds to a negative free-energy curvature in the space of density fluctuations.
It is studied through the $k$-dependent curvature matrix $\mathcal C^f$.
The eigen-mode of minimal curvature is represented in fig.\ref{FIG:Cf-point} 
for one given point of average density $(\rho^0_n,\rho^0_p)$, as a function of $k$, by three quantities : its eigen-value, its direction in the  $(\rho_p,\rho_n)$-plane and  in the $(\rho_p,\rho_e)$-plane. 

The upper part of fig.\ref{FIG:Cf-point} gives the minimal eigen-value $C_<$
of the curvature matrix. 
We observe that, for the chosen average-density point, there is an interval of $k$ 
allowing a negative free-energy curvature in the space of density fluctuations :
the corresponding plane wave will then be amplified  leading to matter clusterization. 
The lower limit of the instability interval is due to the Coulomb term ($\propto 1/k^2$), 
which tends to suppress long wavelenght instabilities which would lead to the 
formation of a macroscopic charge. The upper limit is determined by the nuclear density-gradient term ($\propto k^2$),
which acts against the surface formation associated with short wavelenght fluctuations. 
Thus, both effects arising from finite-size fluctuations lead to a reduction of the instability,
with respect to the case of thermodynamic phase separation in 
uncharged nuclear-matter. 

Let us now consider the phase-separation direction, 
which is given by the eigen-vector associated with the minimal curvature.
The middle part of fig.\ref{FIG:Cf-point} represents this direction projected in the $(\rho_n,\rho_p)$-plane.
Comparing this direction to the line of constant proton fraction, we can see  
that the amplification of fluctuations will lead to a 
dense phase (clusters) which is more symmetric than the surrounding gas. 
This is the well-known phenomenon of isospin distillation 
\cite{Muller-Serot-PRC52}, 
resulting from the minimization of the nuclear symmetry energy.
The lower part of the figure represents the eigen-vector direction in the $(\rho_p,\rho_e)$ plane, 
\emph{i.e.} it represents the mobility of electrons.
This behavior is essentially determined by the relative weight of coulomb interaction and electron incompressibility, as shown by the fact
that the complete neglect of nuclear interaction (dashed line)
changes the results only slightly.
At  low $k$-values, this quantity is dominated by Coulomb energy, 
which forces the electrons to closely follow the protons in order to lower the electric charge. 
Conversely, for high values of $k$, electron incompressibility
dominates over Coulomb interaction, 
and the electron gas tends to stay a uniform background.

In order to better quantify the effect of electrons, 
the phase-separation direction
is compared to two extreme idealized cases represented on the same figure.
In both cases, the proton charge is neutralized by a gas of negative charges :
the incompressibility of this gas is set to either zero or infinity.
In the first case, the Coulomb energy is exactly cancelled : 
the resulting curves show the effect of nuclear density-gradient term, 
the point at $k=0$ corresponding to homogeneous nuclear matter as discussed in section
\ref{SUBSEC:homog-spino}.
In the second case, the negative gas forms a strictly homogeneous background, 
which suppresses the monopolar term of Coulomb interaction.
In reality the electron gas has a finite incompressibility, related to the susceptibility $\chi_e^{-1}$.
As noticed in section 
\ref{SUBSEC:homog-equench},
the high value of $\chi_e^{-1}$ strongly suppresses the electron degree of freedom.
As a result, the electron dimension in the curvature matrix 
(\ref{EQ:Cf})
has a only a perturbative effect with respect to the case of a uniform negative back-ground.
Since the (small) electron mobility allows to lower the Coulomb energy,
it makes the fluctuation more favorable and
leads to a slight increase of the instability interval.

\begin{figure}
\begin{center}
\includegraphics[width=1\linewidth]{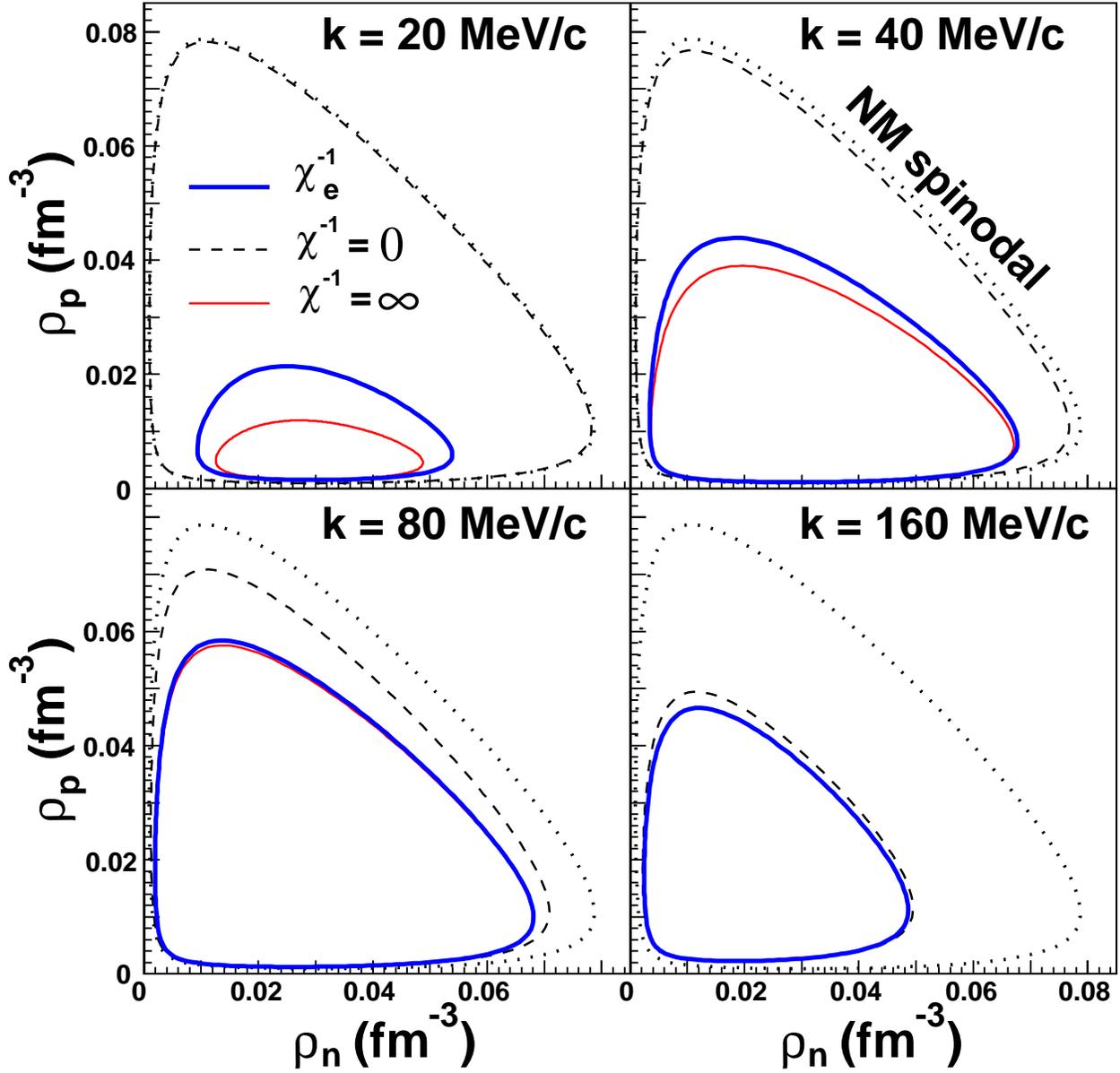}
\caption
	{
	$k$-spinodals for different values of $k$ at $T=5 MeV$.
	The external dotted curve is the thermodynamic spinodal region of nuclear matter, it is 
	the same in all panels.
	The $k$-spinodals of $NMe$ are compared to two limiting cases 
	(zero and infinite incompressibility of the negatively charged gas).
	The thermodynamic spinodal (dotted curve) 
	corresponds to the case $\chi^{-1}=0$ (dashed curve) at $k=0$.
	}
\label{FIG:kspino}
\end{center}
\end{figure}

Fixing different values of $k$, we have represented in fig.\ref{FIG:kspino} 
the corresponding instability regions ($k$-spinodals) 
in the density plane $(\rho_n,\rho_p)$
at a given temperature.
Each point of this plane gives the initial densities of the homogeneous system, 
from which fluctuations start.
The effect of the finite electron incompressibility is more visible 
at lower $k$, where the Coulomb term dominates. 
For a fixed $k$, the perturbation increases as $\chi_e^{-1}$ decreases, 
\emph{i.e.} towards higher values of $\rho^0_e=\rho^0_p$.
Starting
from $k\approx 80$ MeV/c the electron gas can safely be represented
as a fixed background.
%
%
The curves presented on fig.\ref{FIG:kspino} 
can be qualitatively compared with the results of ref.\cite{Providencia-PRC73} 
(obtained in a RMF approach at zero temperature).
Nuclear-matter spinodal region are very similar in both cases.
When finite-size instabilities are considered, 
a similar behaviour is obtain for the higher values of $k$ ($k\gtrsim100$).
For the lower values, Coulomb contribution becomes dominant and is found to affect differently the shape of the $k$-spinodals : fig.\ref{FIG:kspino} shows a stronger reduction of instability in the proton-rich part of the plan, while the curves remain essentially symmetric in the RMF calculation. 
This signs a different evolution of the derivative $\partial_{\rho_n}\mu_n$ through the density plane
(indeed, when Coulomb interaction dominates, the direction of minimal free-energy curvature gets closer to the neutron-axis direction).
%
%

\begin{figure}
\begin{center}
\includegraphics[width=0.6\linewidth]{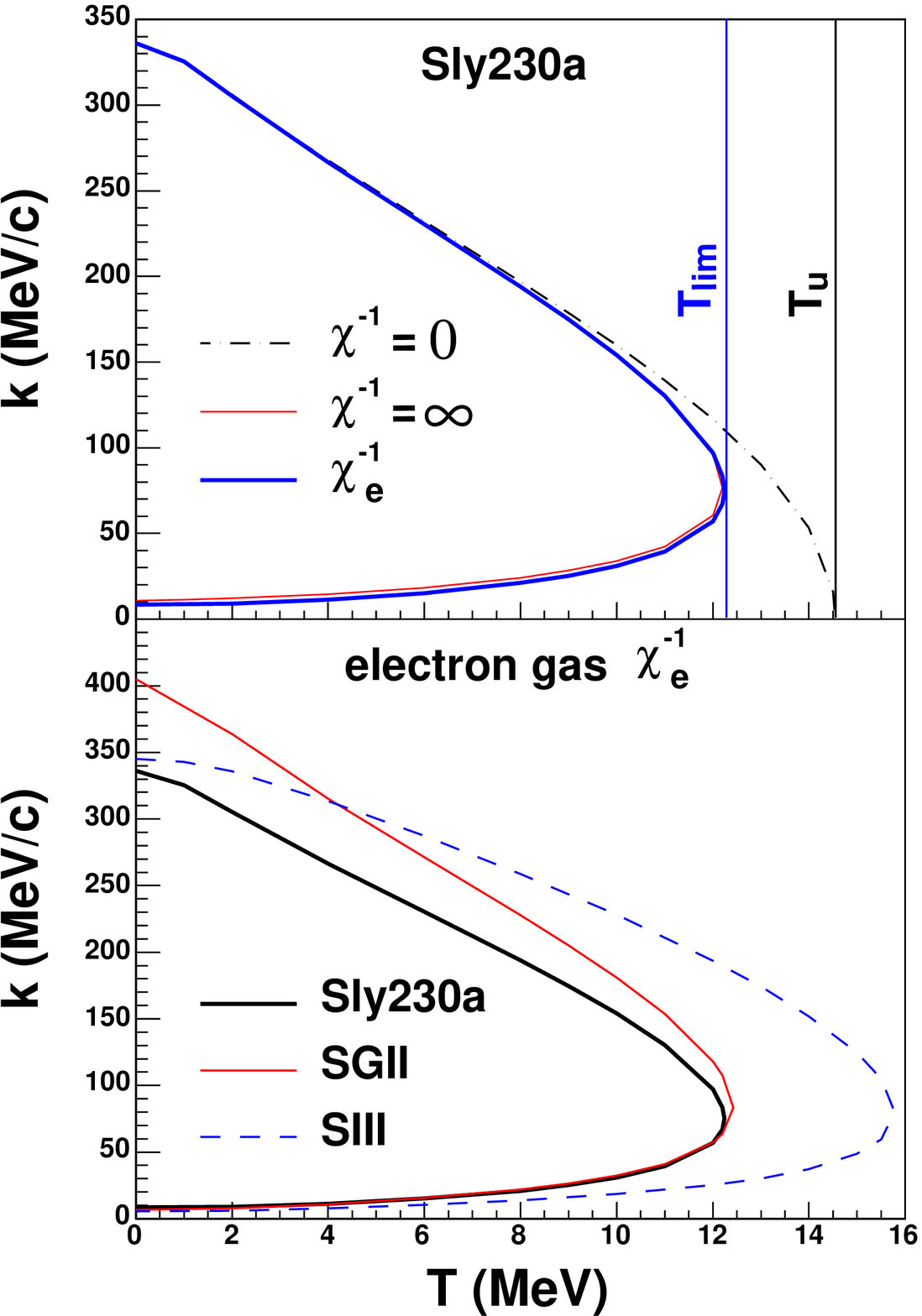}
\caption
	{
	Wave-number $k$ interval for which the density fluctuation
	is amplified in a region of the density plane, as a function of temperature.
	Top : $NMe$ result ($\chi_e^{-1}$, compared to two limiting cases 
	(zero and infinite incompressibility of the negatively charged gas)
	for Sly230a.
	The curve correponding to $\chi^{-1}=0$ goes to zero 
	at the critical temperature of nuclear matter $T_u$.
	For the physical $NMe$ case, the instability interval disappears for a limit
	 temperature
	$T_{lim}<T_u$. This temperature is slightly higher for $NMe$.
	Bottom : $NMe$ curves for three Skyrme parameterizations.
	}
\label{FIG:bornes-k}
\end{center}
\end{figure}

The $k$-interval, for which $NMe$ presents an instability in at least 
one density point,
is represented on fig.\ref{FIG:bornes-k} as a function of the temperature $T$. 
It is maximal at zero temperature, 
and goes to zero at a limiting temperature which is about $12 MeV$ 
for the realistic forces (almost $16 MeV$ for SIII).

\begin{figure}
\begin{center}
\includegraphics[width=1\linewidth]{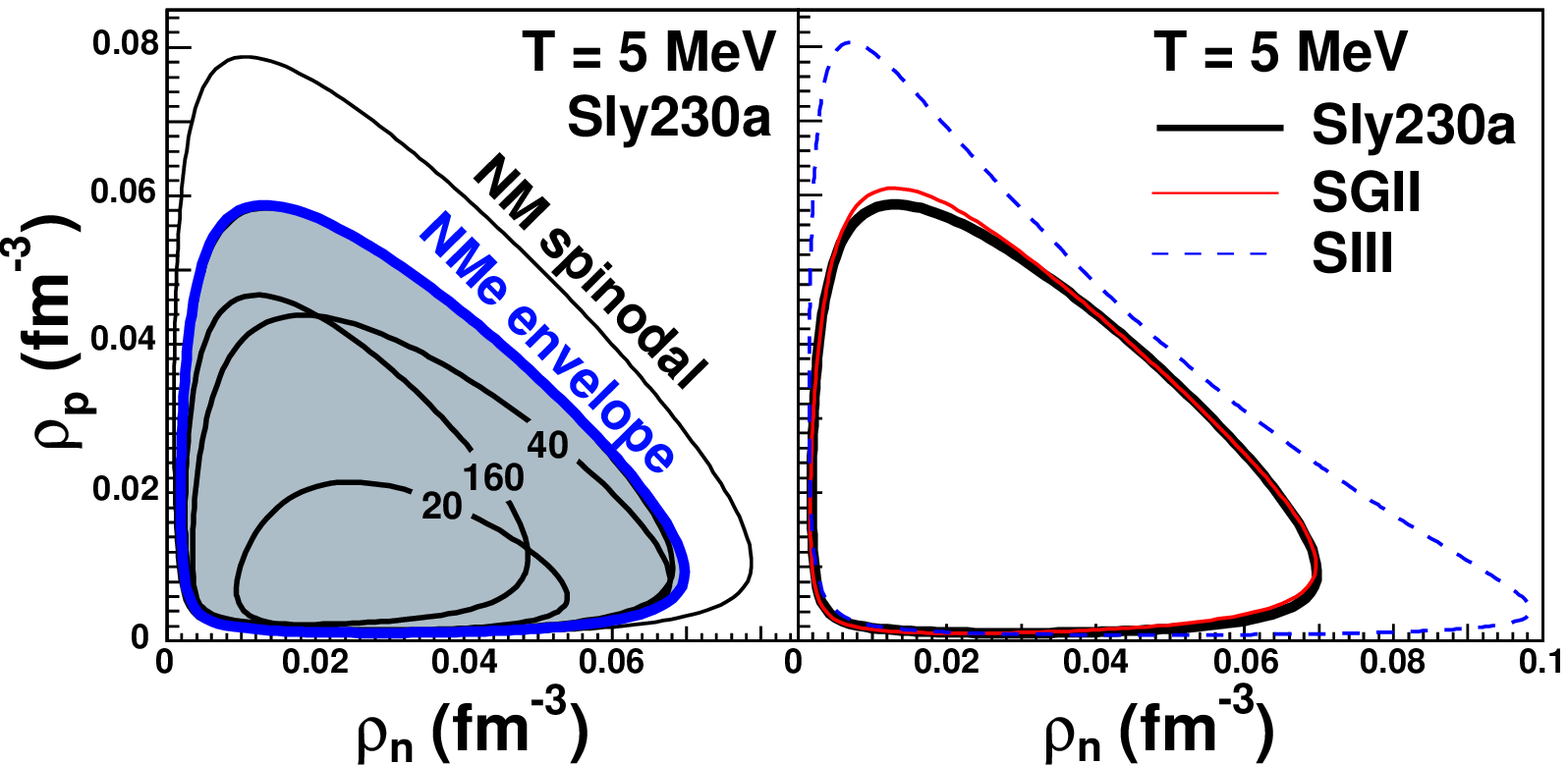}
\caption
	{
	Envelope of instability against finite-size density fluctuations at $T=5 MeV$.
	For every point inside this region, 
	at least one fluctuation mode can be amplified.
	Left : 
	some $NMe$ $k$-spinodals ($k$ values given in $MeV/c$) 
	and the instability envelope (thick curve).
	$NM$ thermodynamic spinodal is also given (external thin curve).
	Right : $NMe$ instability envelope for three different Skyrme forces.
	}
\label{FIG:envelop}
\end{center}
\end{figure}

The global instability region (at a given temperature) 
is defined as the envelope of the different $k$-spinodals : 
it contains all the points presenting an instability against clusterization for at least one value of $k$. 
It is represented in fig.\ref{FIG:envelop} for three different Skyrme parameterizations : 
Sly230a, SGII and SIII. 
We can see that the two realistic forces give similar envelopes. 
SIII presents an atypical behaviour, with a much wider instability region. 
This is essentilly due to the too high value of $K_{\infty}$ (nuclear incompressibility at saturation density) (see Fig.1). 

\subsection{Influence on cluster properties}

In a spinodal-decomposition scenario, 
phase properties are mostly determined by the conditions at the onset of instability
\cite{PC-PhysRep}.
In particular matter clusterization is dominated by the development of the most unstable mode, \emph{i.e.} the wave number giving rise to the fastest  amplification of the fluctuation. The temporal evolution of a fluctuation
of wave-number $k$ is given by:
\begin{equation}
\delta \rho_q = A_q e^{i(\vec k \cdot \vec r - \omega t)} + c.c.,
\end{equation}
where unstable modes are characterized by an imaginary frequence.
Because of the exponential amplification of such modes,
clusterization is dominated by the wave number $k_0$ corresponding 
to the highest imaginary frequence $|\omega \left (k_0\right )|=max$.
The dispersion relation $\omega(k)$ can be established
in a hydrodynamic description. 
This is easily derived
\cite{PC-PhysRep}
in the case of one single fluid, from the Euler and continuity equations:
\begin{eqnarray}
d(m\rho \mathbf{v})/dt = -\nabla P, \\
\partial \rho / \partial t + \nabla (\rho \mathbf{v})=0,
\end{eqnarray}
where $P$ is the system's pressure, related to the free-energy density $f$ by $P=\rho df/d\rho$. 
These equations in the limit of small amplitude fluctuations $\rho=\rho_0+\delta\rho$
can be expressed as a diffusion equation
\begin{equation}
\frac{\partial^2 \rho}{\partial t^2}= \rho_0 C \nabla^2 \rho,
\end{equation}
where $C=d^2f/d\rho^2|_{\rho_0}$ is the curvature 
of the homogeneous-system free-energy density. 
A Fourier representation of this equation leads to the dispersion relation  $\omega^2 =\rho_0 C k^2$.
When this curvature is negative, the frequency is imaginary : 
it can be expressed as $\omega = i/\tau$, where $\tau$ is the time constant
characterizing the development of fluctuations.

%
%

In our study, the instability direction for a given wave-number $k$ 
in each point of the density plane, is associated with the eigen-value $C_< (k)$ 
of the curvature matrix $\mathcal C$. 
Since the temporal evolution of this minimal-curvature mode is given by $\tau^2 \propto 1/(-C_< k^2)$, the most unstable mode for each temperature and density 
can be determined by the condition 
$-C k_0^2=max$.
In the diabatic limit that the successive dynamics can, as a first approximation,
be neglected\cite{PC-PhysRep}, we can link the free-entropy curvature matrix of homogeneous matter, to final cluster properties as issued from the subsequent spinodal decomposition.
In particular, an order of magnitude for cluster sizes is given by the half wavelength $\lambda_0/2 = \pi/k_0$;
cluster isotopic composition is related to the direction of the eigen-vector $\vec u_<(k_0)$ associated with $C_< (k_0)$. 

\begin{figure}
\begin{center}
\includegraphics[width=1\linewidth]{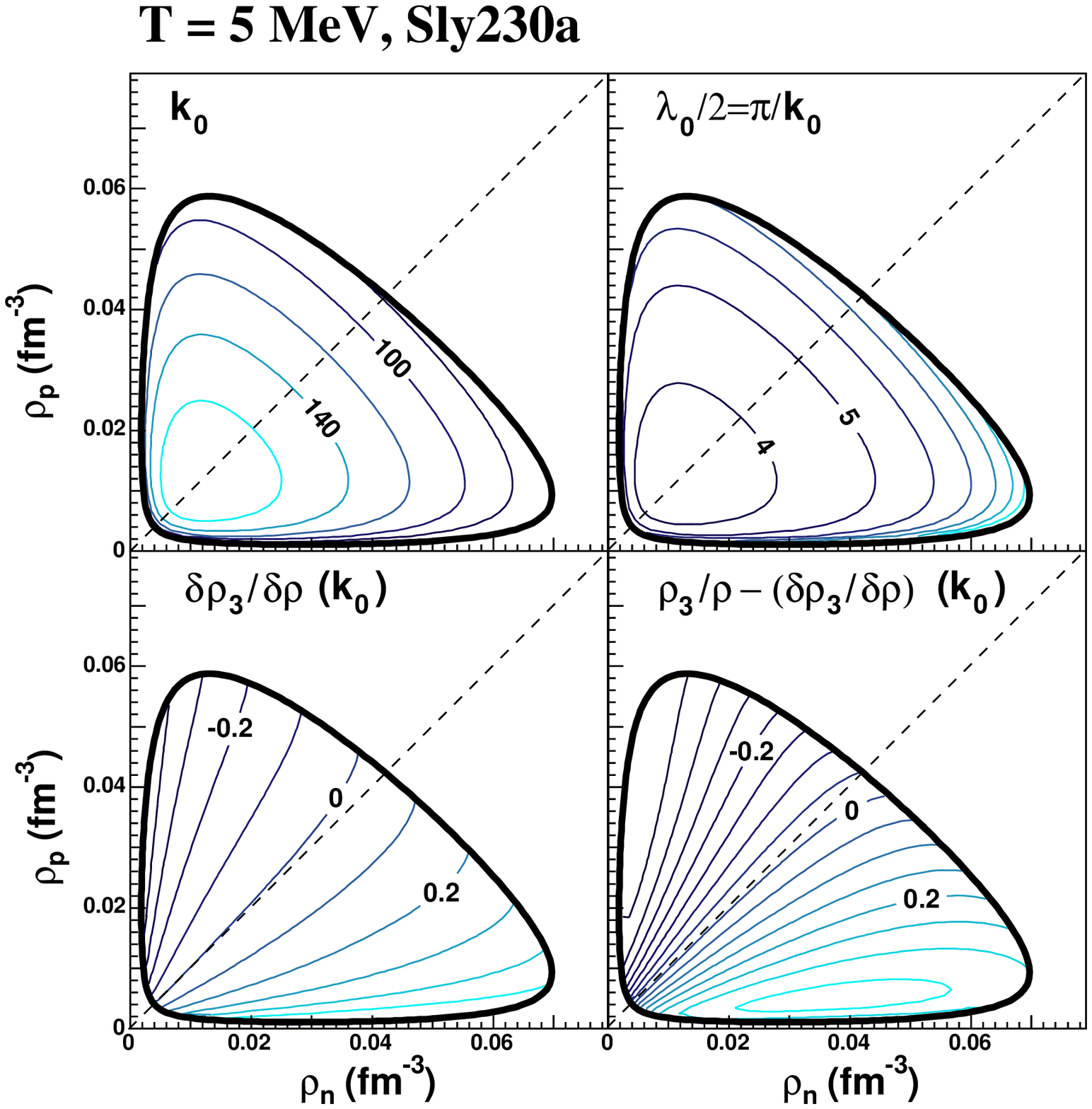}
\caption
	{
	Most unstable mode properties through the density plane, represented in level lines.
	Top-left : 
	most unstable wave number $k_0 (MeV/c)$, 
	levels regularly spaced of $20 MeV/c$.
	Top-right : 
	corresponding half wavelength $\lambda_0/2 (fm)$, 
	levels regularly spaced of $1$$ fm$.
	Bottom-left : 
	most-unstable mode direction in nuclear-density plane $\delta \rho_3/\delta\rho (k_0)$,
	levels regularly spaced of $0.1$.
	Bottom-right : 
	difference respect to the constant $Z/A$ direction, $\rho_3/\rho-\delta \rho_3/\delta\rho (k_0)$, 
	levels regularly spaced of $0.05$.
	}
\label{FIG:cluster-bidim}
\end{center}
\end{figure}

The evolution of the most unstable mode through the density plane 
is given by fig.\ref{FIG:cluster-bidim}, for a fixed temperature. 
The upper part presents quantities linked to cluster size : 
the most unstable wave-number $k_0$, 
and its associated half wavelength $\lambda_0/2$.
For a wide range of asymmetries, the typical cluster size depends essentially on the total density, rising towards the instability boundary.
Level lines circle a minimum at low density, quasi-symmetric matter.
At these low densities the very high frequency of the most unstable mode
gives typical sizes of a few $fm$, corresponding to clusters of $\approx 10$ 
particles. This result should be taken with care: the Thomas-Fermi 
approximation inherent in our treatment breaks down for such small
clusters, and we may be facing the limit of validity
of the mean field approach.

The lower part of the figure presents properties affecting the
isotopic composition of clusters: 
the direction of $\vec u_<(k_0)$ in the nuclear-density plane 
is given by the rapport $\delta \rho_3 / \delta \rho$. 
It is compared to the direction of constant proton fraction by the difference
$\rho_3/\rho - \delta \rho_3 / \delta \rho$.
We can see that clusterization almost systematically 
involves portions of matter which are more symmetric than the global system,
which is again a manifestation of isospin distillation. 

\begin{figure}
\begin{center}
\includegraphics[width=0.8\linewidth]{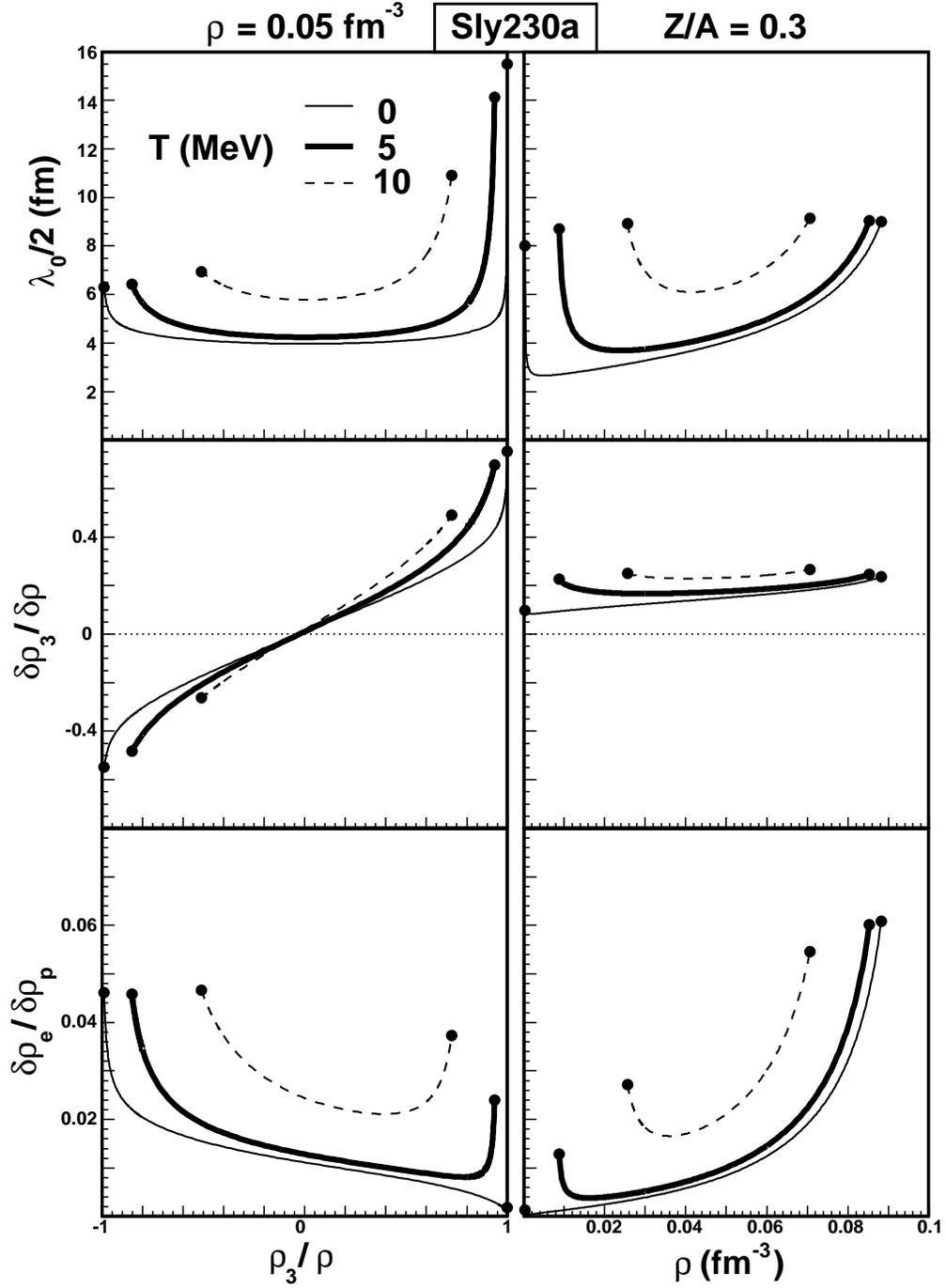}
\caption
	{
	Cluster properties as indicated by the most unstable mode, along two axis :
	one at constant density (left side), the other at constant $Z/A$ (right side).
	Three temperatures are considered with Sly230a.
	}
\label{FIG:cluster-Slya}
\end{center}
\end{figure}

\begin{figure}
\begin{center}
\includegraphics[width=0.8\linewidth]{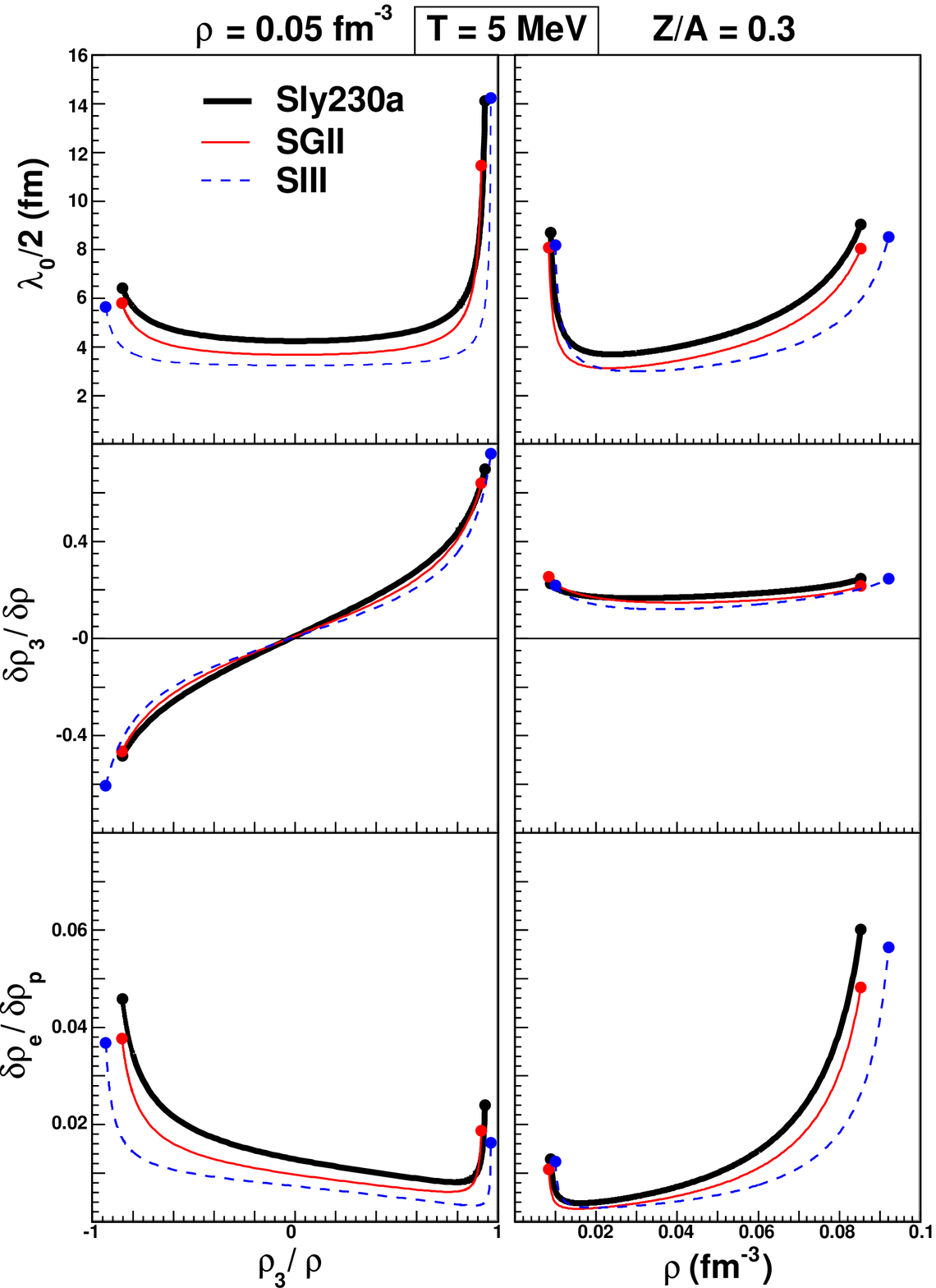}
\caption
	{
	Cluster properties as indicated by the most unstable mode, along two axis :
	one at constant density (left side), the other at constant $Z/A$ (right side).
	Three different Skyrme forces are considered at $T=5 MeV$.
	}
\label{FIG:cluster-T5}
\end{center}
\end{figure}

A representation of the most unstable mode  
is given by fig.\ref{FIG:cluster-Slya} and \ref{FIG:cluster-T5}.
They show some characteristic evolutions for 
the half wavelentgh and the phase separation direction
as a function of density and proton fraction, at different temperatures
and using different Skyrme parametrizations.
We can see that we can expect clusterized stellar matter with 
similar characteristics in a very wide range of temperatures,
densities and asymmetries, and these qualitative results appear 
robust respect to the chosen Skyrme parametrization.
In particular realistic forces, irrespective of their different 
symmetry energy at subsaturation densities, lead to very close results
for all observables. 
Increasing the temperature has the effect of increasing the cluster size
and decreasing the distillation.
The electron participation to phase separation is given by $\delta \rho_e / \delta \rho_p$ : 
this ratio  is low, and correlated with $\lambda_0$ 
(the larger clusters lead to a higher polarization of the electron gas).

\begin{figure}
\begin{center}
\includegraphics[width=0.8\linewidth]{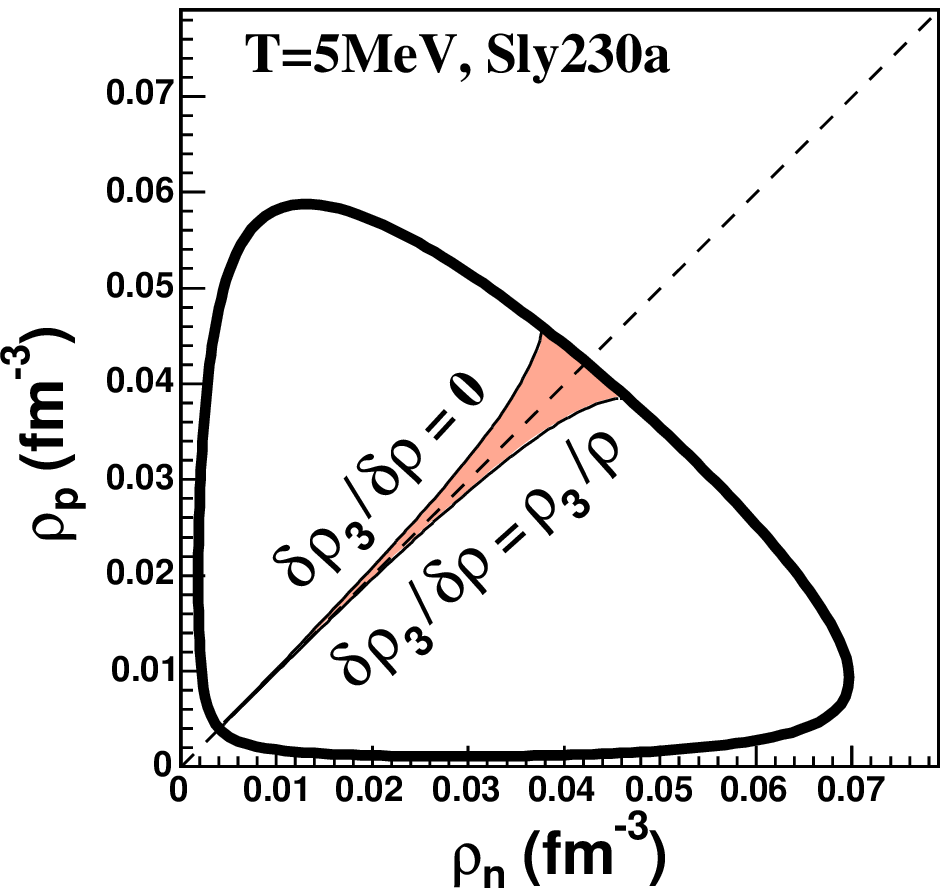}
\caption
	{
	Coexistence region calculated with the Sly230a interaction
	at a temperature $T=5$ MeV. The grey area gives the
	region of abnormal distillation due to the Coulomb interaction,
	situated between the two lines $\delta \rho_3/\delta \rho=0$ 
	and $\delta \rho_3/\delta \rho=\rho_3/\rho$. 
	($\delta \rho_3/\delta \rho$ is the phase-separation direction
	given by the most unstable mode).
	}
\label{FIG:cluster-inv}
\end{center}
\end{figure}

In the absence of Coulomb interaction (as for the study of normal nuclear matter), 
isospin symmetry imposes along the $\rho_n=\rho_p$ axis the relation 
$\delta \rho_3 / \delta \rho=\rho_3/\rho=0$.
In the present case, the presence of a proton charge hinders $\rho_p$ fluctuations, leading to an enhancement of $\delta \rho_3 / \delta \rho$. 
Level-lines $\delta \rho_3 / \delta \rho=0$ and $\rho_3/\rho-\delta \rho_3 / \delta \rho=0$ 
are respectively shifted to the proton-rich and neutron-rich part of the density plane,
delimiting a region of abnormal isospin distillation
, shown in fig.\ref{FIG:cluster-inv} for a temperature $T=5$ MeV.
On the neutron-rich side of this region, 
phase separation leads to a dense phase more asymmetric than the global system, 
and a proton-rich gas. 
On the proton-rich side, for a large enough final density the dense phase is neutron rich.
However, the region concerned by this anomaly is thin : 
for the major part of the instability region, the isospin distillation phenomenon of normal nuclear matter is not qualitatively modified
by the presence of electric charge.

\section{$\beta$-equilibrium in $NMe$} \label{beta}

We now study the case of a $\beta$-equilibrated star matter, 
composed of neutrons, protons, electrons  and neutrinos $\nu$.
For simplicity, leptons will be treated in the degenerated-gas approximation.
If this equilibrium is reached at a given temperature and density, it imposes a given isotopic composition, determined by the rate equality
between reactions of electronic capture and neutron disintegration :
\begin{equation}
\begin{array}{l}
p+e^- \rightarrow n+\nu\\
n \rightarrow p + e^- + \bar{\nu}.
\end{array}
\end{equation}
When the core of a massive star collapses, 
the increase of density displaces $\beta$-equilibrium towards a more neutron-rich composition.

$\beta$-equilibrium corresponds to free-energy minimization.
In the presence of different species $i$ of mass $m_i$, the total 
free energy density is given by :
\begin{equation}
f^t(\{\rho_i\}) = f (\{ \rho_i \}) + \sum_i \rho_i m_i.
\end{equation}
$f^t$ derivatives define the chemical potentials $\mu^t_i$ :
\begin{equation}
\mu^t_i = \partial_{\rho_i} f^t = \partial_{\rho_i}f +m_i = \mu_i +m_i,
\end{equation}
so that $\beta$-equilibrium corresponds to the relation
\begin{equation}
\begin{array}{lll}
\mu^t_p + \mu^t_e & = & \mu^t_n + \mu^t_{\nu} \label{beta_eq}\\ 
\mu_p + m_p + \mu_e + m_e & = & \mu_n + m_n + \mu_{\nu}. \\ 
\end{array}
\end{equation}

\begin{figure}
\begin{center}
\includegraphics[width=1\linewidth]{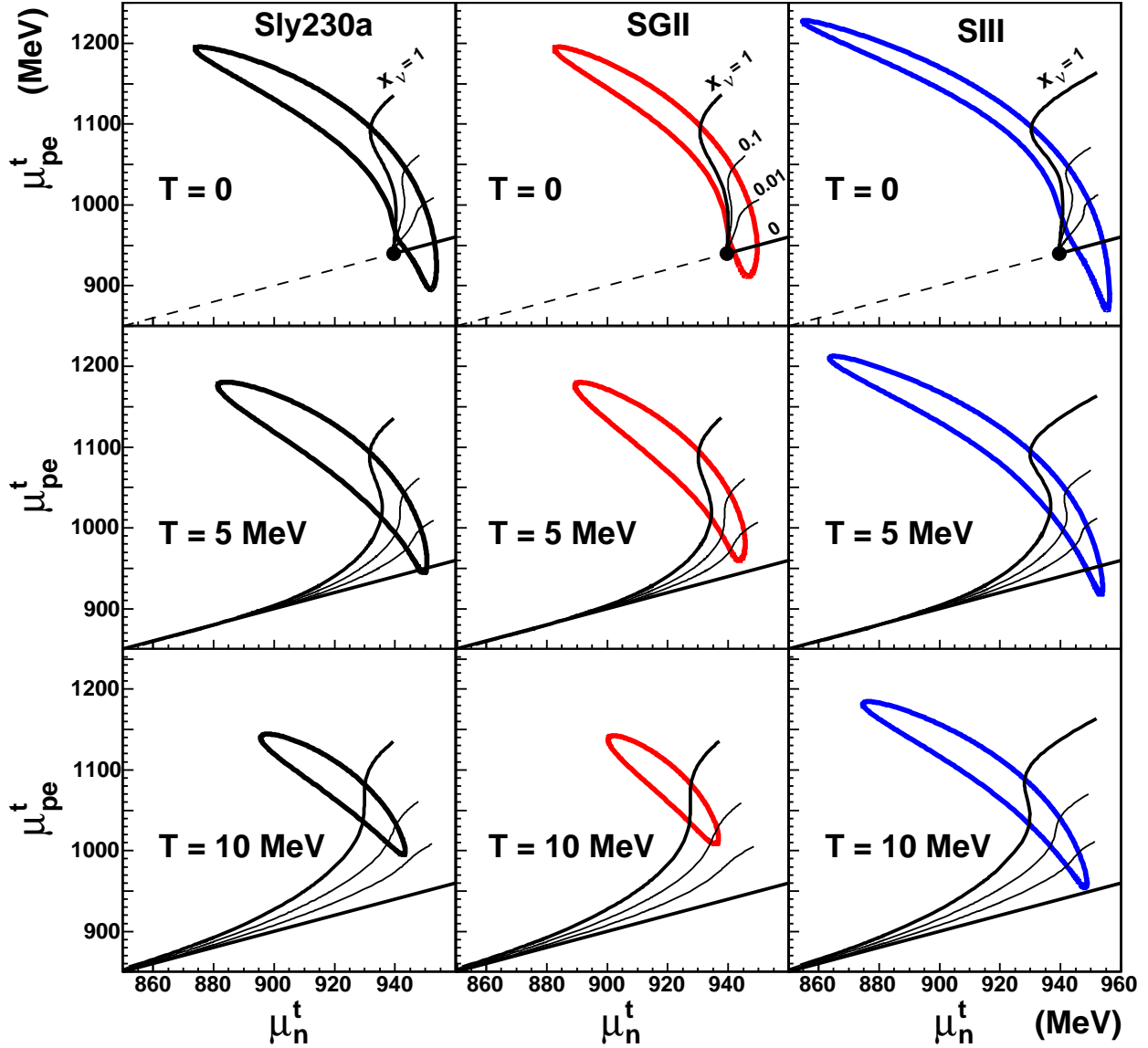}
\caption
	{
	$\beta$-equilibrium in chemical-potential plane $(\mu^t_n,\mu^t_{pe}=\mu^t_p+\mu^t_e)$,
	for different temperatures and neutrino opacities $x_{\nu}=0$, $0.01$, $0.1$ and $1$.
	The lines $x_{\nu}=0$ and $1$ delimit the region 
	which is in principle accessible to $\beta$ equilibrium.
	The $NMe$ instability envelope is also represented.
	}
\label{FIG:eqb-mu}
\end{center}
\end{figure}

Eq.(\ref{beta_eq}) supposes that neutrinos participate in $\beta$-equilibrium.
This is true in the initial post-bounce supernova dynamics, when 
matter is dense enough for neutrinos to be completely trapped in the star.
In the later phase of the cooling process, matter becomes transparent to 
neutrinos \cite{Pons-AJ513, Prakash-LNP578}, 
which then do not equilibrate their chemical potential with 
the one of the other species.
%
%
The transition between these two regimes is smooth, 
and in the following we will study post-bounce supernova matter considering various rates of neutrino trapping.
The detailed relation between density, temperature and neutrino trapping in supernova matter
is not a settled issue, especially concerning the crust region which is of interest for the liquid-gas phase transition. For this reason, for our study we treat the trapping fraction as a free parameter, 
denoted $x_{\nu}$ : it is defined as the total fraction of neutrinos produced by electron-capture reactions which is retained in the star.
%
%
We suppose an initial proton fraction $Z/A_0$ corresponding to $^{56}Fe$ 
(which is the main constituent of the star core before collapse).
For each point $(\rho,Z/A)$ of the nuclear density plane, 
$x_{\nu}$ and $Z/A_0$ determine the neutrino density inside 
the matter according to :
\begin{equation}
\rho_{\nu} = x_{\nu} \rho \left[ (Z/A)_0 - Z/A \right].\label{x_nu}
\end{equation}
%
%
%
Leaving $x_{\nu}$ as a free parameter, we can estimate the effect of neutrino trapping on the thermodynamic properties of the system.
%
%
The neutrino density is linked to the chemical potential 
by the ultra-relativistic degenerate Fermi gas expression :
\begin{equation}
\mu_{\nu} = \hbar c (6 \pi^2 \rho_{\nu})^{1/3}.\label{mu_nu}
\end{equation}
Eqs.(\ref{beta_eq}),(\ref{x_nu}),(\ref{mu_nu}) describe a trajectory 
in the plane of chemical potentials, which fixes the isotopic composition
of matter at each temperature as a function of the 
%
%
$x_{\nu}$ parameter, linked to neutrino opacity.
%
%
The $\beta$-equilibrium compositions corresponding to different values of $x_{\nu}$ are represented in fig.\ref{FIG:eqb-mu} in the space of chemical potentials $(\mu^t_n,\mu^t_{pe}=\mu^t_p+\mu^t_e)$. The diagonal in this
space gives the $\beta$-equilibrium curve for a matter completely
transparent to neutrinos.
Any point of the plane is $\beta$-equilibrated for a value of $\mu_{\nu}$
($i.e.$ a number of trapped neutrinos)
corresponding to the vertical shift respect to this line.
The lines corresponding to $x_{\nu}=0$ and $1$ give the region 
which is in principle accessible to $\beta$-equilibrium : 
this crosses the instability envelope, which is reported as a thick line 
in the chemical-potential plane.
We can see that the matter composition at $\beta$-equilibrium 
is very sensitive to the 
%
%
trapping fraction $x_{\nu}$ :
%
%
only a small fraction of trapped neutrinos is sufficient 
to sensibly increase the stellar proton fraction.  
At zero temperature, the instability envelope is crossed for any value of $x_{\nu}$.
At higher temperatures of the order of $T=10 MeV$, relevant for the initial
dynamical steps of the proto-neutron-star, matter at $\beta$-equilibrium  
is too asymmetric to be unstable against clusterization if all neutrinos
escape from the cooling star, 
while a fraction as small as $x_{\nu}=0.01$ is sufficient 
to lead the system to explore the instability region.
This result depends very slightly on the different Skyrme parametrizations,
meaning that we can reasonably expect that matter will be clusterized
in the whole evolution of the proto-neutron-star.

\begin{figure}
\begin{center}
\includegraphics[width=0.8\linewidth]{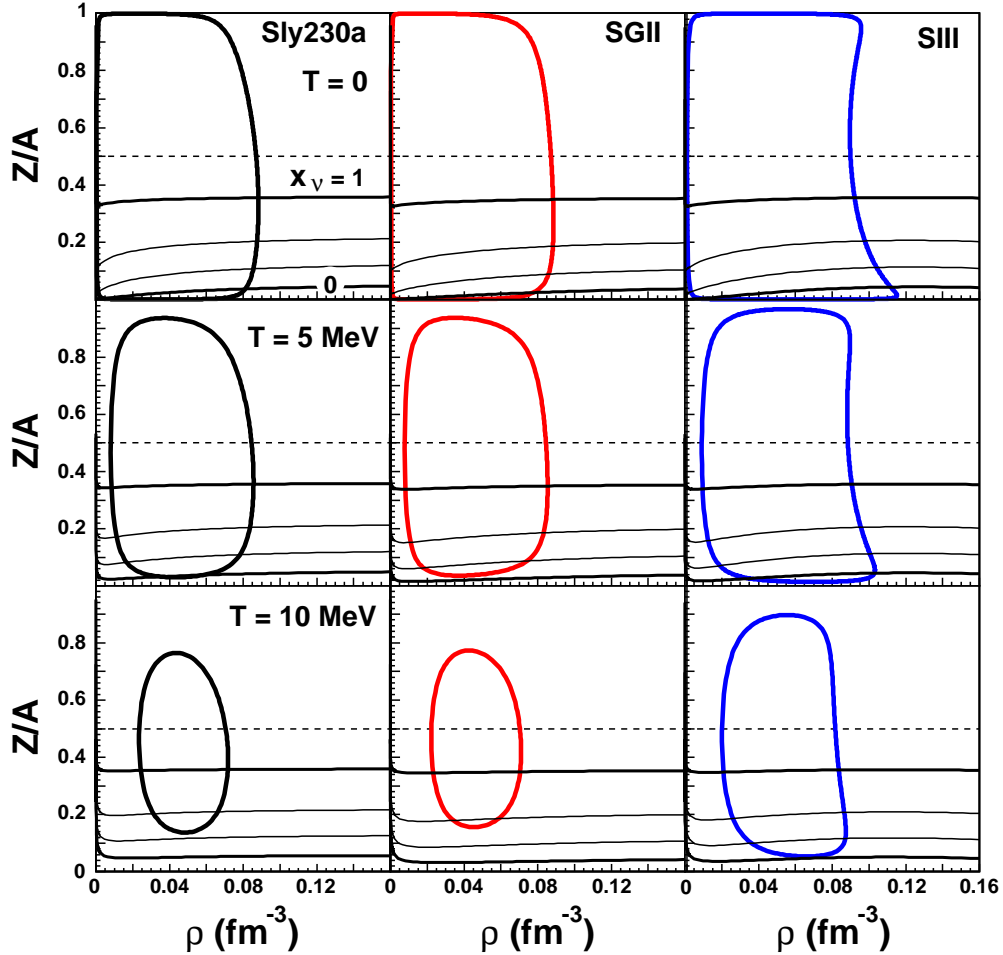}
\caption
	{
	$\beta$-equilibrium in the plane $(\rho,Z/A)$
	for different temperatures and neutrino opacities $x_{\nu}=0$, $0.01$, $0.1$ and $1$.
	$MNe$ instability envelope is also represented.
	}
\label{FIG:eqb-rZsA}
\end{center}
\end{figure}

A representation of $\beta$-equilibrium composition in the plane $(\rho,Z/A)$
is given by fig.\ref{FIG:eqb-rZsA} for different temperatures and Skyrme interactions. 
Each path corresponding to a fixed value of $x_{\nu}$ shows a quasi-constant proton fraction.
The proton fraction shows very little evolution until low densities : 
then it first slightly decreases before going abruptly to $1$ 
in the limit of zero density. 
Indeed, at vanishing density the mass balance $m_n-m_p-m_e>0$ dominates $\beta$-equilibrium, imposing a full decay of neutrons into protons.
However, since this rising happens at very low density, 
fig.\ref{FIG:eqb-rZsA} shows that for a fixed $x_{\nu}$ 
$\beta$-equilibrium essentially corresponds to a constant-$Z/A$ line.

We should stress that in this calculation nuclear matter is always considered as homogeneous.
When clusters are formed, $\beta$-equilibrium does not correspond any more 
to  the simple mean-field trajectories represented in figs.\ref{FIG:eqb-mu} and \ref{FIG:eqb-rZsA}.
Thus the $\beta$-equilibrium curves displayed by these figures represent the initial condition for matter clusterization.
The instability envelope we have represented 
corresponds to the region where clusters can be formed by spinodal decomposition.
This gives a lower-limit estimate for the density interval corresponding to 
clusterization.
Indeed the region for which completely equilibrated matter is clusterized 
has to be wider than the spinodal-instability envelope.
However, outside this envelope, 
matter can in principle be homogeneous and $\beta$-equilibrated
since in a situation of meta-stability 
the time needed for cluster formation can be longer than $\beta$-equilibrium time scale.
Let us also consider the case the case of a thermally equilibrated initially homogenous dense matter.
If it is submitted to a fast dilatation, $Z/A$ remains constant while density decreases.
Then, the proportion of neutrinos trapped in the initial state, which fixes the constant composition $Z/A$, 
determines wether the instability envelope is crossed or not.


\begin{figure}
\begin{center}
\includegraphics[width=0.8\linewidth]{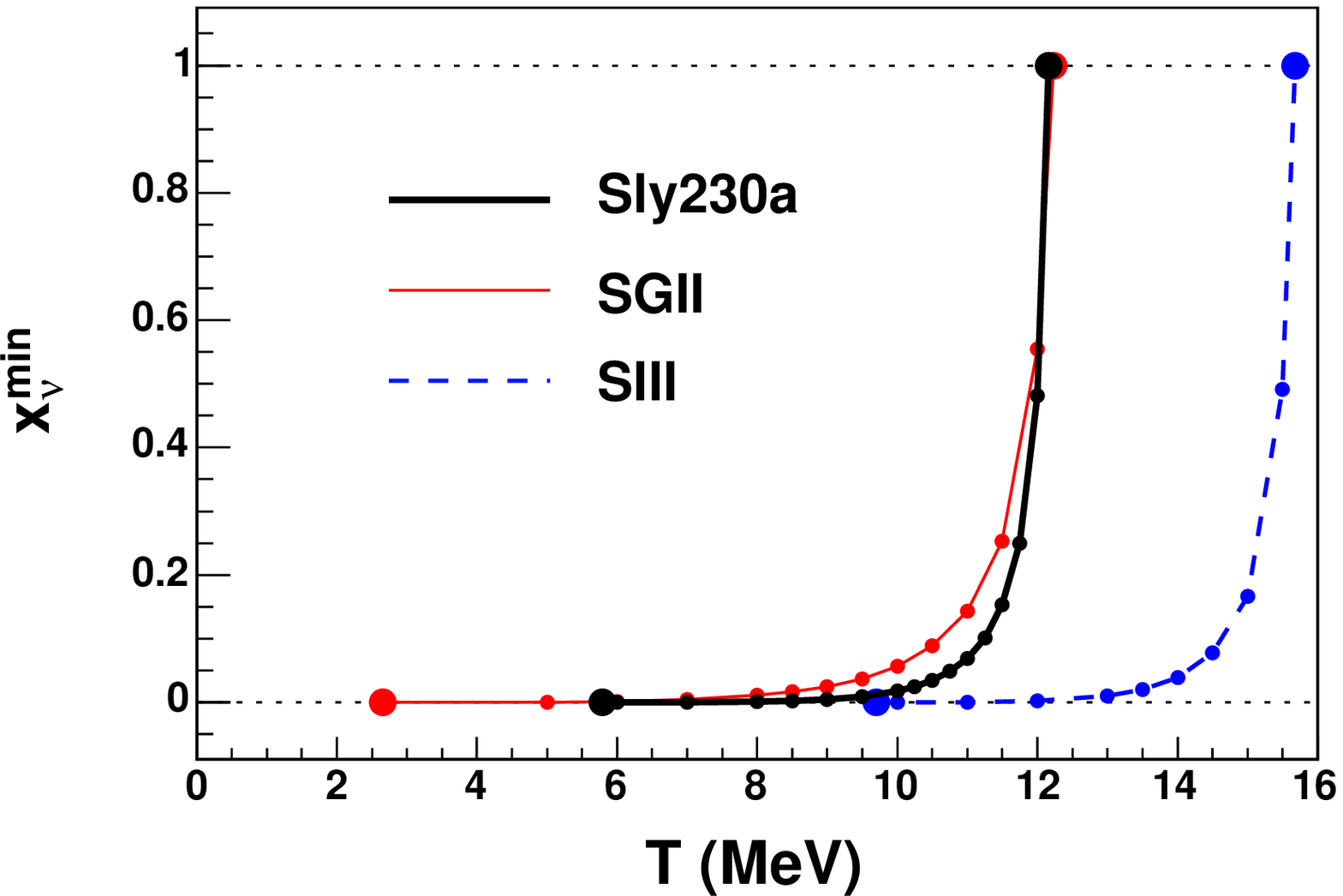}
\caption
	{
	Temperature dependance of the minimum neutrino opacity $x_{\nu}^{min}$
	leading to the instability region on a path of $\beta$-equilibrium.
	}
\label{FIG:eqb-xlim}
\end{center}
\end{figure}

Figure \ref{FIG:eqb-xlim} gives a quantitative view of the neutrino fraction
needed for $\beta$-equilibrated matter 
to reach the region of instability against clusterization,  
as a function of temperature.
The value $x_{\nu}^{min}$ corresponds to the path at constant $x_{\nu}$ 
which is tangential to the instability envelope.
The curves obtained show a slow evolution 
with very small values of neutrino fraction at low temperature, 
and a sharp enhancement as the limiting temperature for instability is approached.
This profile is due to the typical evolution of the instability region, 
since, as we have seen, $\beta$-equilibrium paths depend only weakly on the temperature.

\section{Summary and conclusion}

The nuclear-matter liquid-gas phase transition results in a region of instability 
against cluster formation in low-density finite-temperature star matter composed of neutrons, protons and electrons.
This has been studied analyzing homogeneous-matter instabilities 
against a plane-wave density fluctuation, in the small-amplitude limit.
In a spinodal decomposition scenario, cluster properties as sizes and 
chemical compositions
can be predicted from the characteristics of the most unstable mode
at the onset of fluctuation amplification :
they are determined studying the free-energy curvature matrix in the space of density fluctuations,
as a function of the wave number $k$.

For a cold neutron star, thermodynamic equilibrium is reached.
The liquid-gas transition concerns the inner crust, where pasta phases are expected.
Our approach does not include the study of such structures, 
since it is based on homogeneous-matter properties : 
however, the instability envelope which we obtain gives the minimal region
where equilibrated matter is clusterized.

The out-of-equilibrium phase separation may concern 
proto-neutron stars issuing from core-collapse in supernova explosions,
since the shock-wave causes abrupt changes in the density.
At this time, star matter is hot and contains trapped neutrinos, 
whose propagation is a key issue for the understanding of supernova dynamics
\cite{Buras-PRL90}.
It has been pointed out\cite{Horowitz-PRC69} that neutrino in-medium propagation is affected by the eventual presence of clusters.
On the other hand, we have found that presence of neutrinos affects the star-matter composition in a significant way. Indeed neutrino trapping rules the crossing of the instability region, which in turn leads to the formation
of clusters.
This interplay is sensitive to the different time scales implied in the explosion :
temperature and density evolution, 
as well as the realization of $\beta$ and thermodynamic equilibrium.


\end{document}